\documentclass[aps,prl,preprint,groupedaddress]{revtex4-1}

\usepackage{graphicx}

\usepackage{amsmath}

\usepackage{color}

\usepackage{epstopdf}

\begin{document}

\title{Magnetic Field Generation by the Rayleigh-Taylor Instability \\in Laser-Driven Planar Plastic Targets}

\author{L. Gao$^{1,2}$, P. M. Nilson$^{1,3}$, I. V. Igumenschev$^{1}$, S. X. Hu$^{1}$, J. R. Davies$^{1,2,3}$, C. Stoeckl$^{1}$, \\M. G. Haines$^{4}$, D. H. Froula$^{1}$, R. Betti$^{1,2,3,5}$, and D. D. Meyerhofer$^{1,2,3,5}$}

\affiliation{$^{1}$Laboratory for Laser Energetics, University of Rochester, Rochester, NY, 14623, USA} 
\affiliation{$^{2}$Department of Mechanical Engineering, University of Rochester, Rochester, NY, 14623, USA}
\affiliation{$^{3}$Fusion Science Center for Extreme States of Matter, University of Rochester, Rochester, NY, 14623, USA}
\affiliation{$^{4}$Department of Physics, Imperial College, London SW7 2AZ United Kingdom}
\affiliation{$^{5}$Department of Physics and Astronomy, University of Rochester, 
Rochester, NY, 14623, USA}

\date{\today}

%---------------------------------------------------

\begin{abstract}

\noindent Magnetic fields generated by the Rayleigh-Taylor instability were measured in laser-accelerated planar foils using ultrafast proton radiography. Thin plastic foils were irradiated with $\sim$4-kJ, 2.5-ns laser pulses focused to an intensity of $\sim$10$^{14}$ W$/$cm$^{2}$ on the OMEGA EP Laser System. Target modulations were seeded by laser nonuniformities and amplified during target acceleration by the Rayleigh-Taylor instability. The experimental data show the hydrodynamic evolution of the target and MG-level magnetic fields generated in the broken foil. The experimental data are in good agreement with predictions from 2-D magnetohydrodynamic simulations.

\end{abstract}

%---------------------------------------------------

%\pacs{}

%---------------------------------------------------

\maketitle

%---------------------------------------------------

Target designs predicted to achieve ignition by inertial confinement fusion (ICF) rely on understanding the Rayleigh-Taylor (RT) instability \cite{Rayleigh1883,Taylor22031950,Betti2006}. When an ICF capsule is imploded, the ablation front during the acceleration phase and the pusher-fuel interface during the deceleration and stagnation phase are RT unstable \cite{lindl1995,McCrory}. At the unstable interface, spikes of higher-density plasma penetrate into lower-density plasma and bubbles of lower-density plasma rise through the higher-density plasma. Understanding the RT instability is important because it can amplify capsule perturbations and destroy implosion uniformity.

Previous theoretical work showed that a plasma subject to RT instability should generate spontaneous magnetic fields \cite{Stamper1971,Mima1978}. These fields may exist in inertial fusion plasmas and modify electron thermal transport \cite{Braginskii1965,Haines1997}. If present and unaccounted for, these fields may degrade implosion performance compared to theoretical predictions \cite{Stamper1991,Evans1986,Srinivasan2012}. Magnetic fields can be generated in high-energy-density plasmas by many different mechanisms \cite{Haines1986}, including the thermoelectric effect \cite{StamperRipin1975,Colombant1977}, anisotropic hot-electron velocity distributions \cite{Estabrook1978}, and thermoelectric instability \cite{Haines1981}. Recently the first measurement of RT-induced magnetic fields was reported \cite{Manuel2012}. This work showed RT-induced magnetic fields in laser-accelerated targets with preimposed target-surface modulations from experiments on the OMEGA Laser Facility \cite{Boehly1997}. Magnetic fields with strengths of up to 0.1 MG were inferred in the linear growth phase of the RT instability using face-on mono-energetic proton radiography \cite{Li2006}. The mono-energetic protons were generated from D-$^{3}$He fusion inside an imploding capsule.

This Letter reports magnetic field generation during the nonlinear growth phase of the RT instability in an ablatively-driven plasma using ultrafast laser-driven proton radiography \cite{Borghesi_PRL_2004}. Thin plastic foils were irradiated with $\sim$4-kJ, 2.5-ns laser pulses focused to $\sim$10$^{14}$ W$/$cm$^{2}$ on the OMEGA EP Laser System at the University of Rochester's Laboratory for Laser Energetics \cite{Waxer2005}. The driven foils were probed with an ultrafast proton beam that revealed the magnetohydrodynamic (MHD) evolution of the target. The target modulations were seeded by laser nonuniformities and amplified during the target-acceleration phase. These experiments show, for the first time, MG-level magnetic fields inside a laser-driven foil broken apart by the RT instability. The experimental results are consistent with 2-D MHD calculations using the code {\it DRACO} \cite{Keller1999,igumenshchev2009}. 

\begin{figure}[t]
\includegraphics[width=8cm]{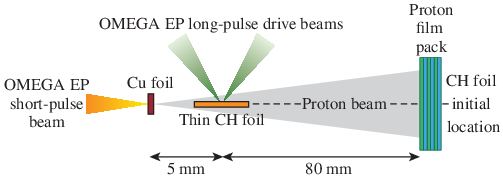}
\caption{\label{fig1} Experimental setup.}
\end{figure}

Figure \ref{fig1} shows a schematic of the experimental setup. Two long-pulse beams irradiated a 15- or 25-$\mu$m-thick CH foil. The foil area was 5 $\times$ 5 mm$^{2}$. Only a central $\sim$1-mm-diam part of the foil was driven. Each laser beam delivered an $\sim$2-kJ pulse with a wavelength of 351 nm and a 2.5-ns square temporal profile at 23$^{\circ}$ to the target normal. The laser beams were focused to $\sim$850-$\mu$m-diam focal spots using distributed phase plates \cite{DPP_LLEReview}. The average overlapped intensity was $\sim$4 $\times$ 10$^{14}$ W$/$cm$^{2}$. 

The CH foil was probed in a direction orthogonal to the main interaction with an ultrafast proton beam \cite{Clark2000,Snavely2000}. The proton source was generated by irradiating a planar, 20-$\mu$m-thick Cu foil with an $\sim$1-kJ, 10-ps pulse at a wavelength of 1.054 $\mu$m \cite{wilks2001}. The laser pulse was focused with a 1-m-focal-length, $f$/2 off-axis parabolic mirror onto the Cu foil at normal incidence, providing an intensity of $\sim$5 $\times$ 10$^{18}$ W$/$cm$^{2}$. The relative timing between the long-pulse and the short-pulse beams was measured with an x-ray streak camera. Protons were accelerated from the surface of the Cu foil to tens of MeV by target normal sheath acceleration (TNSA) \cite{wilks2001}. The TNSA mechanism generated a highly laminar proton beam with a micron-scale virtual source size \cite{Cowan2004}, providing high spatial resolution for probing the main interaction with point-projection radiography \cite{Borghesi_PRL_2004}.
 
Combining a filtered stack detector with time-of-flight dispersion provided a multiframe imaging capability \cite{Mackinnon2006}. The high-energy protons that passed through the driven CH target were detected with a stack of radiochromic film interleaved with aluminum filters. Soft x rays were filtered with an additional aluminum foil on the front surface of the stack. Each film layer recorded a different probe time because the transit time for protons to the CH foil varied with energy. Protons with different energies deposited energy inside various film layers corresponding to their energy-dependent Bragg peak. The temporal coverage obtained in these experiments on a single shot was $\sim$120 ps, with spatial and temporal resolutions of $\sim$5 to 10 $\mu$m and $\sim$10 ps, respectively. The image magnification $M$ $=$ $\left( L + l \right)/l$, where $l$ is the distance from the proton-source foil to the CH target and $L$ is the distance from the CH target to the radiochromic film detector. For these experiments, $M$ was $\sim$17 to 20, depending on the radiochromic film layer.

Figure 2 shows a typical proton radiograph of a 25-$\mu$m-thick CH foil unbroken by instability formation. This radiograph was obtained with 13-MeV protons at time {\it t} $=$ {\it t}$_{0}$ $+$ 2.56 ns, where {\it t}$_{0}$ is the arrival time of the long-pulse beams at the target surface. The undriven foil horizon is indicated. The long-pulse beams irradiated the target from the left and the blowoff plasma accelerated the central part of the foil toward the right. The driven foil had a transverse size comparable with the laser focal spot. At this time, the foil had a velocity of (3$\pm$1) $\times$ 10$^{7}$ cm$/$s, calculated from the measured driven-foil trajectory history. 

Thinner-foil targets were broken by instability formation during the acceleration phase. Figure 3 shows proton radiographs for a 15-$\mu$m-thick CH foil driven with the same laser conditions as the 25-$\mu$m-thick foil case. These data were obtained with 13-MeV protons from two different shots. The relative timing with respect to {\it t}$_{0}$ was varied from 2.11 ns to 2.56 ns. At {\it t} $=$ {\it t}$_{0}$ $+$ 2.56 ns, the foil has traveled a greater distance than the 25-$\mu$m-thick foil because less mass was accelerated. In this case, bubble-like structures are observed in the proton radiographs. These perturbations grow in time and show that the target has broken apart during the acceleration phase. Larger-scale structures at {\it t} $=$ {\it t}$_{0}$ $+$ 2.56 ns indicate this growth.

\begin{figure}[t]
\includegraphics[width=6cm]{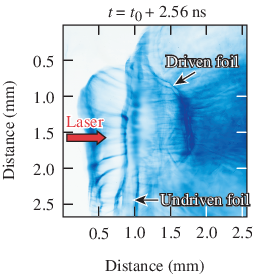}
\label{fig2}
\caption{ Proton radiograph of a 25-$\mu$m-thick CH foil taken with 13-MeV protons at {\it t} $=$ {\it t}$_{0}$ $+$ 2.56 ns. The laser drive, the foil horizon, and the bow-shaped driven foil are indicated.}
\end{figure}
 
Further evidence for the broken foil is provided by the appearance of plasma beyond the driven target. Figure 3 shows a plasma sheath ahead of the RT-unstable region. Hot plasma in the laser-ablation region has fed through the compromised foil and formed a halo around the unstable expanding matter. A sheath electric field forms at the plasma-vacuum interface and is detected in the proton radiographs. This effect is not observed in the radiographs of the stable, 25-$\mu$m-thick foil, uncompromised by instability growth (Figure 2).

\begin{figure}[t]
\includegraphics[width=6cm]{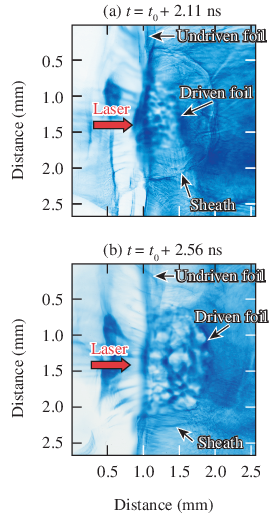}
\label{fig3} 
\caption{Proton radiographs of a 15-$\mu$m-thick CH foil taken with 13-MeV protons at (a) {\it t} $=$ {\it t}$_{0}$ $+$ 2.11 ns and (b) {\it t} $=$ {\it t}$_{0}$ $+$ 2.56 ns, where {\it t}$_{0}$ is the arrival time of the long-pulse beams at the CH-foil surface. The laser drive, the foil horizon, the RT-unstable plasma, and the sheath field formed by hot-plasma feedthrough are indicated.}
\end{figure}

The main observation from these data is the electromagnetic fields that are generated during the RT instability growth. In proton radiography, proton beam density modulations are caused by deflections from electromagnetic fields and by collisional scattering and stopping inside the probed target. For these experiments, collisional scattering and proton stopping are small. For example, collisional energy losses for 13-MeV protons passing through $\sim$30-$\mu$m-thick solid CH are {\it $\Delta$E}$/${\it E} $<$ 1$\%$. Electromagnetic fields must play a dominant role in generating the bubble-like structures observed in the radiography data. The broken foil is revealed in the data by electromagnetic fields that are generated at the RT-unstable interface.

\begin{figure}[t]
\includegraphics[width=9cm]{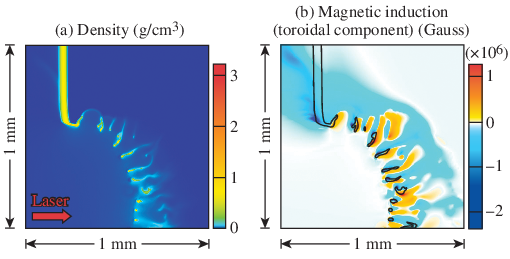}
\label{fig4}
\caption{(a) Simulated density profile at {\it t} $=$ {\it t}$_{0}$ $+$ 2.1 ns. The modeled target is axisymmetric about the horizontal axis. (b) Self-generated magnetic field distribution at {\it t} $=$ {\it t}$_{0}$ $+$ 2.1 ns. The density contour for $\rho$ $=$ 0.05 g$/$cm$^{3}$ is overlaid.}
\end{figure}

This interpretation is supported by numerical modeling with the 2-D resistive MHD code {\it DRACO} \cite{Keller1999,igumenshchev2009}. {\it DRACO} has a 2-D cylindrical geometry. The equation governing the magnetic field is

\begin{equation}
\frac{\partial \mathbf{B}} {\partial t} = \nabla \times (\mathbf{V} \times \mathbf{B}) + \frac{c}{e}\left [ \nabla \times \left(\frac{\nabla p_{e}}{n_{e}}\right)-\nabla \times \frac{\left (\nabla \times \mathbf{B}\right )\times \mathbf{B}}{4\pi n_{e}}- \nabla\times\frac{\mathbf{R_{T}}+ \mathbf{R_{u}}}{n_{e}} \right ]
\end{equation}

\mbox{}

\noindent where $\mathbf{B}$ is the magnetic induction, $p_{e}$ is the electron pressure, $n_{e}$ is the electron number density, $e$ is the fundamental unit of charge, $\mathbf{V}$ is the flow velocity, and $\mathbf{R_{T}}$ and $\mathbf{R_{u}}$ are the thermal and frictional forces \cite{Braginskii1965}, respectively. The second term on the right hand side of equation (1) is the thermoelectric source term that is driven by non-parallel density and temperature gradients. The nonuniform $\nabla p_{e}$ force induces poloidal current loops that wrap around the magnetic field toroids. The full Braginskii transport coefficients \cite                 {Braginskii1965}, including the Nernst term \cite{nishiguchi1985} and cross-gradient thermal fluxes, were used in calculating $\mathbf{R_{T}}$ and $\mathbf{R_{u}}$. The temporal evolution of the laser power was provided by experimental measurements. The growth of RT instability in the calculations was seeded by assuming a pre-imposed surface perturbation with a 50-$\mu$m wavelength and a 1-$\mu$m peak-to-valley amplitude. This mode grew fastest when the simulation was initialized without pre-imposed modulations and the final perturbations were developed from numerical noise. 
 
The {\it DRACO} calculations show a 15-$\mu$m-thick foil broken apart by the RT instability, generating MG-level magnetic fields at the RT-unstable interface. Figure 4(a) shows the calculated target-density profile at {\it t} $=$ {\it t}$_{0}$ $+$ 2.1 ns. Density perturbations have grown by the RT instability that are greater in extent than the target thickness, breaking the foil apart. Large density and temperature gradients form in this unstable plasma and spontaneously generate MG-level magnetic fields. Figure 4(b) shows the predicted magnetic field distribution at {\it t} $=$ {\it t}$_{0}$ $+$ 2.1 ns. Overlaid on this field distribution is the calculated density contour for $\rho$ $=$ 0.05 g$/$cm$^{3}$, indicating the position of the target. Magnetic fields generated at the ablation surface are convected towards the lower density corona by the ablated plasma and to higher density regions by hot electrons that carry the heat flux (the Nernst effect) \cite{nishiguchi1985}. In our case, the Nernst convection significantly overperforms the convection by the ablation flow. Up to 2 MG magnetic fields are observed beyond the coronal plasma and inside the driven foil.

{\it DRACO} simulations show that the dynamic effect of the spontaneous magnetic fields on the RT instability is negligible in the linear and the moderately nonlinear stages of its evolution. The fields begin to enhance the RT growth in the highly nonlinear stages when the spike sizes are comparable to, and larger than, the perturbation wavelengths. The {\it DRACO} calculations reproduce the measured foil velocity to within experimental error, indicating that the gross hydrodynamics of the driven foil are as predicted. For a 25-$\mu$m-thick target, {\it DRACO} calculations show that the RT instability does not break the foil apart and no significant small-scale magnetic fields are generated.

Estimates for the magnitude of the generated magnetic fields are made by measuring the angular deflection $\theta$ of protons from their original trajectory while passing through the field region. When the apparent displacement of protons is $\delta$ in the target plane, the angular deflection $\theta$ is calculated by $\tan \theta =  M \delta /D $, where $M$ is the geometric magnification and $D$ is the distance between the main target and the radiochromic film detector. The proton-path-integrated B field caused by the Lorentz force acting upon the proton probe beam is $\int \mathbf{B} \times d\mathbf{l} = m_p v \sin \theta /e$, where $m_{p}$ is the proton mass and $v$ is the proton speed. In our experiments, the protons are deflected by azimuthal magnetic fields generated around the RT spikes. $\delta$ is the radius of a bubble. At {\it t} $=$ {\it t}$_{0}$ $+$ 2.11 ns, a $\delta$ of 25 $\mu$m results in a deflection angle $\theta$ of 0.31$^{\circ}$. Assuming an integration path length slightly larger than the target thickness ($L$ $\sim$ 25 $\mu$m) gives a magnetic field strength of $\sim$1.4 MG, which is in good agreement with the {\it DRACO} simulations.

At the RT-unstable interface, narrow spikes are formed where the dense matter falls through the light matter, and bubbles are generated when the light material rises into the dense material \cite{Betti2006}. This process generates magnetic fields wrapping around the troughs of the spikes. The growth of the spatial scale length of the perturbed features is caused by magnetic field evolution as the RT instability develops. The magnetic field topology in {\it DRACO} is different from the real 3-D situation. In 3-D RT instability, magnetic fields are formed around single spikes and bubbles. However, the magnitude and the predominant wavelength of the magnetic fields are expected to be accurate.
 
\begin{figure}[t]
\includegraphics[width=6cm]{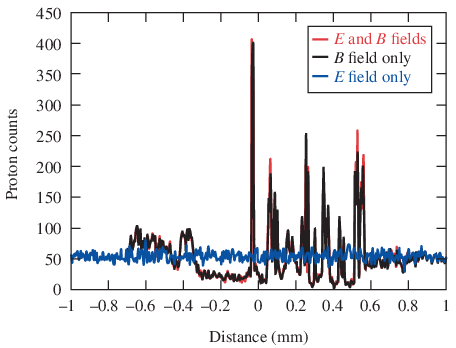}
\label{fig5}
\caption{Proton tracking code results. Proton deflections are modeled based on electromagnetic field distributions predicted by 2-D $DRACO$ calculations.}
\end{figure}

A proton ray tracing code using electromagnetic field distributions from the 2-D $DRACO$ calculations supports the dominant role of magnetic fields in deflecting protons in these experiments. The initial proton-source details and the radiography geometry were taken from the experiments. The accumulated proton numbers were monitored in the ray tracing code at a simulated detector plane. Figure 5 shows the effect of electric and magnetic fields in this process. The predicted proton distribution is unchanged when electric fields are turned off in the calculations, while few proton deflections are observed when magnetic fields are turned off. Self-generated magnetic fields at the RT-unstable interface are the dominant cause for proton-beam deflections in these experiments. Two-dimensional Fourier analysis of the measured proton radiographs shows that the characteristic spatial scale length of the bubble-like features at  {\it t} $=$ {\it t}$_{0}$ $+$ 2.11 ns is $\sim$82 $\mu$m, growing to $\sim$115 to 230 $\mu$m at  {\it t} $=$ {\it t}$_{0}$ $+$ 2.56 ns. Broadly consistent with this experimental trend, Fourier analysis of the proton distribution in Fig. 5 gives a characteristic spatial scale length of $\sim$93 $\mu$m, growing to $\sim$220 $\mu$m at the latest time.

In summary, magnetic field generation during the nonlinear growth of target perturbations by the RT instability in ablatively-driven foils was studied. Measurements of MG-level magnetic fields were supported by recovering characteristic spatial scale lengths of the proton deflections using a particle ray tracing code that incorporates electromagnetic field distributions from a 2-D MHD model. Electric fields were found to be negligible compared to the generated magnetic fields in producing the modulated patterns in the proton radiography beam profile. Simulations suggest that the dynamic effect of these magnetic fields on the RT growth is not significant.

This work was supported by the U.S. Department of Energy Office of Inertial Confinement Fusion under Cooperative Agreement No. DE-FC52-08NA28302, the University of Rochester, and the New York State Energy Research and Development Authority. The support of DOE does not constitute an endorsement by DOE of the views expressed in this article.

%---------------------------------------------------
\bibliographystyle{unsrt}
\bibliography{Manuscript_071012}

\begin{thebibliography}{10}

\bibitem{Rayleigh1883}
Lord Rayleigh.
\newblock Investigation of the character of the equilibrium of an
  incompressible heavy fluid of variable density.
\newblock {\em Proc. London Math. Soc.}, 14:170, 1883.

\bibitem{Taylor22031950}
Geoffrey Taylor.
\newblock The instability of liquid surfaces when accelerated in a direction
  perpendicular to their planes.
\newblock {\em Proceedings of the Royal Society of London. Series A.
  Mathematical and Physical Sciences}, 201(1065):192--196, 1950.

\bibitem{Betti2006}
R.~Betti and J.~Sanz.
\newblock Bubble acceleration in the ablative \mbox{R}ayleigh-\mbox{T}aylor
  instability.
\newblock {\em Phys. Rev. Lett.}, 97:205002, Nov 2006.

\bibitem{lindl1995}
John Lindl.
\newblock Development of the indirect-drive approach to inertial confinement
  fusion and the target physics basis for ignition and gain.
\newblock {\em Physics of Plasmas}, 2(11):3933--4024, 1995.

\bibitem{McCrory}
{R. L. McCrory}, {D. D. Meyerhofer}, {S.J. Loucks}, {S. Skupsky}, {R. Betti},
  {T. R. Boehly}, {T. J. B. Collins}, {R. S. Craxton}, {J. A. Delettrez}, {D.
  H. Edgell}, {R. Epstein}, {K. A. Fletcher}, {C. Freeman}, {J. A. Frenje}, {V.
  Yu. Glebov}, {V. N. Goncharov}, {D. R. Harding}, {I. V. Igumenshchev}, {R. L.
  Keck}, {J. D. Kilkenny}, {J. P. Knauer}, {C. K. Li}, {J. Marciante}, {J. A.
  Marozas}, {F. J. Marshall}, {A. V. Maximov}, {P. W. McKenty}, {S. F. B.
  Morse}, {J. Myatt}, {S. Padalino}, {R. D. Petrasso}, {P. B. Radha}, {S. P.
  Regan}, {T. C. Sangster}, {F. H. S\'eguin}, {W. Seka}, {V. A. Smalyuk}, {J.
  M. Soures}, {C. Stoeckl}, {B. Yaakobi}, and {J. D. Zuegel}.
\newblock Progress in direct-drive inertial confinement fusion research at the
  \mbox{L}aboratory for \mbox{L}aser \mbox{E}nergetics.
\newblock {\em J. Phys. IV France}, 133:59--65, 2006.

\bibitem{Stamper1971}
J.~A. Stamper, K.~Papadopoulos, R.~N. Sudan, S.~O. Dean, E.~A. McLean, and
  J.~M. Dawson.
\newblock Spontaneous magnetic fields in laser-produced plasmas.
\newblock {\em Phys. Rev. Lett.}, 26:1012--1015, Apr 1971.

\bibitem{Mima1978}
K.~Mima, T.~Tajima, and J.~N. Leboeuf.
\newblock Magnetic field generation by the \mbox{R}ayleigh-\mbox{T}aylor
  instability.
\newblock {\em Phys. Rev. Lett.}, 41:1715--1719, Dec 1978.

\bibitem{Braginskii1965}
S.~I. Braginskii.
\newblock {\em Review of Plasma Physics}.
\newblock Consultant Bureau, New York, 1965.

\bibitem{Haines1997}
M.~G. Haines.
\newblock Saturation mechanisms for the generated magnetic field in nonuniform
  laser-matter irradiation.
\newblock {\em Phys. Rev. Lett.}, 78:254--257, Jan 1997.

\bibitem{Stamper1991}
J.~A. Stamper.
\newblock Review on spontaneous magnetic fields in laser-produced plasmas:
  Phenomena and measurements.
\newblock {\em Laser and Particle Beams}, 9(04):841--862, 1991.

\bibitem{Evans1986}
R.~G. Evans.
\newblock The influence of self-generated magnetic fields on the
  \mbox{R}ayleigh-\mbox{T}aylor instability.
\newblock {\em Plasma Physics and Controlled Fusion}, 28(7):1021, 1986.

\bibitem{Srinivasan2012}
Bhuvana Srinivasan, Guy Dimonte, and Xian-Zhu Tang.
\newblock Magnetic field generation in \mbox{R}ayleigh-\mbox{T}aylor unstable
  inertial confinement fusion plasmas.
\newblock {\em Phys. Rev. Lett.}, 108:165002, Apr 2012.

\bibitem{Haines1986}
M.~G. Haines.
\newblock Magnetic-field generation in laser fusion and hot-electron transport.
\newblock {\em Canadian Journal of Physics}, 64(8):912--919, 1986.

\bibitem{StamperRipin1975}
J.~A. Stamper and B.~H. Ripin.
\newblock Faraday-rotation measurements of megagauss magnetic fields in
  laser-produced plasmas.
\newblock {\em Phys. Rev. Lett.}, 34:138--141, Jan 1975.

\bibitem{Colombant1977}
D.~G. Colombant and N.~K. Winsor.
\newblock Thermal-force terms and self-generated magnetic fields in
  laser-produced plasmas.
\newblock {\em Phys. Rev. Lett.}, 38:697--701, Mar 1977.

\bibitem{Estabrook1978}
Kent Estabrook.
\newblock Qualitative aspects of underdense magnetic fields in laser-fusion
  plasmas.
\newblock {\em Phys. Rev. Lett.}, 41:1808--1811, Dec 1978.

\bibitem{Haines1981}
M.~G. Haines.
\newblock Thermal instability and magnetic field generated by large heat flow
  in a plasma, especially under laser-fusion conditions.
\newblock {\em Phys. Rev. Lett.}, 47:917--920, Sep 1981.

\bibitem{Manuel2012}
M.~J.-E. Manuel, C.~K. Li, F.~H. S\'eguin, J.~Frenje, D.~T. Casey, R.~D.
  Petrasso, S.~X. Hu, R.~Betti, J.~D. Hager, D.~D. Meyerhofer, and V.~A.
  Smalyuk.
\newblock First measurements of rayleigh-taylor-induced magnetic fields in
  laser-produced plasmas.
\newblock {\em Phys. Rev. Lett.}, 108:255006, Jun 2012.

\bibitem{Boehly1997}
T.~R Boehly, D.~L Brown, R.S Craxton, R.~L Keck, J.~P Knauer, J.~H Kelly, T.~J
  Kessler, S.~A Kumpan, S.~J Loucks, S.~A Letzring, F.~J Marshall, R.~L
  McCrory, S.~F.~B Morse, W~Seka, J.~M Soures, and C.~P Verdon.
\newblock Initial performance results of the \mbox{OMEGA} \mbox{l}aser
  \mbox{s}ystem.
\newblock {\em Optics Communications}, 133(1–6):495 -- 506, 1997.

\bibitem{Li2006}
C.~K. Li, F.~H. S\'eguin, J.~A. Frenje, J.~R. Rygg, R.~D. Petrasso, R.~P.~J.
  Town, P.~A. Amendt, S.~P. Hatchett, O.~L. Landen, A.~J. Mackinnon, P.~K.
  Patel, V.~A. Smalyuk, T.~C. Sangster, and J.~P. Knauer.
\newblock Measuring \mbox{E} and \mbox{B} fields in laser-produced plasmas with
  monoenergetic proton radiography.
\newblock {\em Phys. Rev. Lett.}, 97:135003, Sep 2006.

\bibitem{Borghesi_PRL_2004}
M.~Borghesi, A.~J. Mackinnon, D.~H. Campbell, D.~G. Hicks, S.~Kar, P.~K. Patel,
  D.~Price, L.~Romagnani, A.~Schiavi, and O.~Willi.
\newblock Multi-\mbox{M}e\mbox{V} proton source investigations in ultraintense
  laser-foil interactions.
\newblock {\em Phys. Rev. Lett.}, 92:055003, Feb 2004.

\bibitem{Waxer2005}
L.~J. Waxer, T.~J.~Kessler D.~N.~Maywar, J. H.~Kelly, R.~L.~McCrory
  B.~E.~Kruschwitz, S. J.~Loucks, C.~Stoeckl D.~D.~Meyerhofer, S. F. B.~Morse,
  and J.~D. Zuegel.
\newblock High-energy petawatt capability for the \mbox{O}mega \mbox{L}aser.
\newblock {\em Opt. Photon. News}, 16(7):30--36, Jul 2005.

\bibitem{igumenshchev2009}
I.~V. Igumenshchev, F.~J. Marshall, J.~A. Marozas, V.~A. Smalyuk, R.~Epstein,
  V.~N. Goncharov, T.~J.~B. Collins, T.~C. Sangster, and S.~Skupsky.
\newblock The effects of target mounts in direct-drive implosions on
  \mbox{OMEGA}.
\newblock {\em Physics of Plasmas}, 16(8):082701, 2009.

\bibitem{DPP_LLEReview}
Laboratory for Laser~Energetics.
\newblock {\em LLE Review}, 33:1--10, 1987.

\bibitem{Clark2000}
E.~L. Clark, K.~Krushelnick, M.~Zepf, F.~N. Beg, M.~Tatarakis, A.~Machacek,
  M.~I.~K. Santala, I.~Watts, P.~A. Norreys, and A.~E. Dangor.
\newblock Energetic heavy-ion and proton generation from ultraintense
  laser-plasma interactions with solids.
\newblock {\em Phys. Rev. Lett.}, 85:1654--1657, Aug 2000.

\bibitem{Snavely2000}
R.~A. Snavely, M.~H. Key, S.~P. Hatchett, T.~E. Cowan, M.~Roth, T.~W. Phillips,
  M.~A. Stoyer, E.~A. Henry, T.~C. Sangster, M.~S. Singh, S.~C. Wilks,
  A.~MacKinnon, A.~Offenberger, D.~M. Pennington, K.~Yasuike, A.~B. Langdon,
  B.~F. Lasinski, J.~Johnson, M.~D. Perry, and E.~M. Campbell.
\newblock Intense high-energy proton beams from petawatt-laser irradiation of
  solids.
\newblock {\em Phys. Rev. Lett.}, 85:2945--2948, Oct 2000.

\bibitem{wilks2001}
S.~C. Wilks, A.~B. Langdon, T.~E. Cowan, M.~Roth, M.~Singh, S.~Hatchett, M.~H.
  Key, D.~Pennington, A.~MacKinnon, and R.~A. Snavely.
\newblock Energetic proton generation in ultra-intense laser--solid
  interactions.
\newblock {\em Physics of Plasmas}, 8(2):542--549, 2001.

\bibitem{Cowan2004}
T.~E. Cowan, J.~Fuchs, H.~Ruhl, A.~Kemp, P.~Audebert, M.~Roth, R.~Stephens,
  I.~Barton, A.~Blazevic, E.~Brambrink, J.~Cobble, J.~Fern\'andez, J.-C.
  Gauthier, M.~Geissel, M.~Hegelich, J.~Kaae, S.~Karsch, G.~P. Le~Sage,
  S.~Letzring, M.~Manclossi, S.~Meyroneinc, A.~Newkirk, H.~P\'epin, and
  N.~Renard-LeGalloudec.
\newblock Ultralow emittance, multi-\mbox{M}e\mbox{V} proton beams from a laser
  virtual-cathode plasma accelerator.
\newblock {\em Phys. Rev. Lett.}, 92:204801, May 2004.

\bibitem{Mackinnon2006}
A.~J. Mackinnon, P.~K. Patel, M.~Borghesi, R.~C. Clarke, R.~R. Freeman,
  H.~Habara, S.~P. Hatchett, D.~Hey, D.~G. Hicks, S.~Kar, M.~H. Key, J.~A.
  King, K.~Lancaster, D.~Neely, A.~Nikkro, P.~A. Norreys, M.~M. Notley, T.~W.
  Phillips, L.~Romagnani, R.~A. Snavely, R.~B. Stephens, and R.~P.~J. Town.
\newblock Proton radiography of a laser-driven implosion.
\newblock {\em Phys. Rev. Lett.}, 97:045001, Jul 2006.

\bibitem{nishiguchi1985}
A.~Nishiguchi, T.~Yabe, and M.~G. Haines.
\newblock Nernst effect in laser-produced plasmas.
\newblock {\em Physics of Fluids}, 28(12):3683--3690, 1985.

\end{thebibliography}
%---------------------------------------------------

\end{document}